\documentclass[12pt]{article}
\usepackage[T1]{fontenc}
\usepackage[latin1]{inputenc}
\usepackage{amsmath}

\makeatletter
\providecommand{\LyX}{L\kern-.1667em\lower.25em\hbox{Y}\kern-.125emX\@}

\makeatother
\input{tcilatex}
\begin{document}

\title{The one-dimensional $XXZ$ model with long-range interactions}
\author{L. L. Gonçalves$^{1}$, A. P. Vieira$^{2}$ and L. P. S. Coutinho$^{1}$}
\maketitle

\begin{abstract}
The one-dimensional $XXZ $ model ($s=1/2 $, $N $ sites) with uniform
long-range interactions among the transverse components of the spins is
considered. The Hamiltonian of the model is explicitly given by $H=J\sum
^{N}_{j=1}\left( s^{x}_{j}s^{x}_{j+1}+s^{y}_{j}s^{y}_{j+1}\right) -(I/N)\sum
^{N}_{j,k=1}s^{z}_{j}s^{z}_{k}-h\sum ^{N}_{j=1}s^{z}_{j}, $ where the $%
s^{x,y,z} $ are half the Pauli spin matrices. The model is exactly solved by
applying the Jordan-Wigner fermionization, followed by a Gaussian
transformation. In the absence of the long-range interactions ($I=0 $), the
model, which reduces to the isotropic $XY $ model, is known to exhibit a
second-order quantum phase transition driven by the field at zero
temperature. It is shown that in the presence of the long-range interactions
($I\ne 0 $) the nature of the transition is strongly affected. For $I>0 $,
which favours the ordering of the transverse components of the spins, the
transition is changed from second- to first-order, due to the competition
between transverse and $xy $ couplings. On the other hand, for $I<0 $, which
induces complete frustration of the spins, a second-order transition is
still present, although the system is driven out of its usual universality
class, and its critical exponents assume typical mean-field values.
\end{abstract}

{\raggedright \textbf{$^{1}$}Departamento de Física da UFC, Campus do Pici,
Cx. Postal 6030, 60451-970 Fortaleza (CE), Brazil \vspace{1.4em} \noindent}

$^{2}$Instituto de Física, Universidade de São Paulo, Cx. Postal 66318,
05315-970 São Paulo (SP), Brazil

\smallskip The observed critical behaviour of magnetic materials, in the
very low temperature limit, has renewed the interest in the study of 
magnetic quantum transitions (S.L.Sondhi,S.M.Girvin,J.P.Carini and D.Shahar,
Rev. Mod. Phys. \textbf{69, 315(1997)) . }Since\textbf{\ }these transitions,
which are governed by quantum fluctuations, occur at T=0, the
one-dimensional models play a important role in their study . Therefore, we
will consider the exactly soluble one-dimensional $XXZ$ model ($s=1/2$),
with a uniform long-range interaction among the spins along the $z$
direction. Due to the long-range interaction, the model also presents
classical critical behaviour, with transitions of first and second order,
and  it has already been considered by Suzuki (J. Phys. Soc. Jpn. \textbf{21}%
, 2140 (1966)). Since  his study was restricted to the analysis of the
classical second order transition of the model, and  we are interested in
its quantum transitions, the model  will be considered again. In particular,
we will be interested in the effect of the long-range interaction on its
quantum critical behaviour.

The Hamiltonian  of the model is given by 
\begin{equation}
H=J\sum_{j=1}^{N}(s_{j}^{x}s_{j+1}^{x}+s_{j}^{y}s_{j+1}^{y})-\frac{I}{N}%
\sum_{j,k=1}^{N}s_{j}^{z}s_{k}^{z}-h\sum_{j=1}^{N}s_{j}^{z},
\end{equation}
where $J>0,N$ is the number of sites of the lattice  and we assume periodic
boundary conditions. By applying the Jordan-Wigner fermionization\cite
{jordan28, lieb61}, followed by a Gaussian transformation, we can write the
partition function of the model as 
\begin{equation}
Z_{N}=\mbox {Tr}e^{-\beta H}=C(\beta )\int e^{-\frac{N}{2}z^{2}}\mbox {Tr}%
e^{-\tilde{H}(z)}dz,
\end{equation}
with 
\begin{equation}
\tilde{H}(z)=\frac{\beta J}{2}\sum_{j=1}^{N}(c_{j}^{\dagger
}c_{j+1}+c_{j+1}^{\dagger }c_{j})-\tilde{h}(z)\sum_{j=1}^{N}c_{j}^{\dagger
}c_{j},
\end{equation}
where $\tilde{h}(z)=\beta (h-I)+\sqrt{2\beta I}z$, $C(\beta )$ depends only
on the temperature, a boundary term has been neglected in $\tilde{H}(z)$,
and the $c_{j}$ are fermion operators.

Introducing the Fourier transforms 
\begin{equation}
c_{j}=\frac{1}{\sqrt{N}}\sum e^{-ikj}\hat{c}_{k},
\end{equation}
we can rewrite $\tilde{H}(z) $ in the diagonal form 
\begin{equation}
\tilde{H}(z)=\sum _{k}\tilde{\varepsilon }_{k}(z)\hat{c}^{\dagger }_{k}\hat{c%
}_{k},
\end{equation}
where $\tilde{\varepsilon }_{k}(z)=\beta J\cos k-\tilde{h}(z) $ and, due to
the periodic boundary conditions, $k=2\pi n/N $ ($n=1,\ldots ,N) $. The
partition function is then given by 
\begin{equation}  \label{partfunc}
Z_{N}=C(\beta )\int e^{-\frac{N}{2}z^{2}}\prod _{k}\left[ 1+e^{-\tilde{%
\varepsilon }_{k}(z)}\right] dz,
\end{equation}
which, in the thermodynamic limit ($N\rightarrow \infty $), can be evaluated
by the saddle-point method. By explicit calculation, we conclude that 
\begin{equation}
m\equiv \left\langle \frac{1}{N}\sum ^{N}_{j=1}s^{z}_{j}\right\rangle =\frac{%
z_{0}}{\sqrt{2\beta I}}-\frac{1}{2},
\end{equation}
where $z_{0} $ is the value of $z $ which makes the integrand in Eq. (\ref
{partfunc}) a maximum.

Noticing that $z_{0}/\sqrt{2\beta I}$ is just the average number of fermions
per energy level, we can write the equation of state of the system, 
\begin{equation}
m=\frac{1}{2\pi }\int_{0}^{2\pi }\frac{dk}{1+e^{-\tilde{\varepsilon}_{k}(m)}}%
-\frac{1}{2},  \label{mzt}
\end{equation}
where $\tilde{\varepsilon}_{k}(m)=\beta J\cos k+\beta (h+2Im)$. In the limit 
$T\rightarrow 0$ ($\beta \rightarrow \infty $), for $(h+2Im)/\leq J,$ Eq. (%
\ref{mzt}) takes the form 
\begin{equation}
m+\tfrac{1}{2}=\frac{1}{\pi }\arccos \left( -\frac{h+2Im}{J}\right) ,
\label{mz0}
\end{equation}
which, for $I=0$, readily reduces to the well-known expression for the $XX$
chain \cite{niemeijer67}. \ To analyze the behavior of the model near the
quantum critical point, assuming $h\ge 0$, we define the order parameter 
\cite{delima94} $\sigma \equiv \frac{1}{2}-m$ and expand Eq. (\ref{mz0}) to
second order in $\sigma \rightarrow 0^{+}$, obtaining 
\begin{equation}
\frac{\pi ^{2}}{2}\sigma ^{2}-\frac{2I}{J}\sigma =\frac{h_{c}-h}{J}\ge 0,
\label{sig0}
\end{equation}
where $h_{c}=J-I$. For $I=0$ we regain the usual $XX$ chain result 
\begin{equation}
\sigma \sim (h_{c}-h)^{1/2},
\end{equation}
while for $I<0$ we get the expected mean-field scaling form 
\begin{equation}
\sigma \sim (h_{c}-h)^{1}.
\end{equation}
Note that Eq. (\ref{sig0}) cannot be satisfied for $I>0$, an indication that
in this case the model undergoes a first-order transition at $h=h_{c}$ to a
state where the transverse magnetization is saturated ($m=1/2$). In this
case, there is a hysteresis cycle associated to the transition, which is due
to the presence of the metastable states. These states can be identified by
looking at the free energy which, for $(h+2Im)/\leq J$, and in the limit $%
T\rightarrow 0$, is explicitly given by

\begin{equation}
f=\frac{h}{2}-\frac{J}{\pi }\left[ \sin \varphi +\left( \frac{h+2Im}{J}%
\right) \varphi \right] +I(m^{2}+m),  \label{m1}
\end{equation}
where $\varphi $ is defined as 
\begin{equation}
\varphi =\arccos \left( -\frac{h+2Im}{J}\right) .
\end{equation}
Taking the limit $h\rightarrow 0$ in Eq. (\ref{m1}), and by imposing that $%
f(0)=f(1/2),$ which are minima of the free energy, we can show that the
systems presents spontaneous magnetization for $I/J\ge 4/\pi $ .

The previous analysis allows us to determine the phase diagram of the model
at zero temperature, shown in Figure 1. Notice that there must be a finite
temperature critical line ending at the point $(h/J,I/J)=(1,0) $, which is
thus analogous to a bicritical point. The finite temperature behavior of the
model will be considered in future work.

\textbf{Acknowledments.} This work was partially financed by the Brazilian
agencies CNPq, FINEP and Fapesp. A. P. Vieira thanks T. A. S. Haddad and S.
R. Salinas for useful discussions.

\vspace{0.3cm}

\newpage

\textbf{Figure caption.}

Figure 1. Phase diagram of the model at $T=0 $. The solid and dashed lines
indicate second- and first-order phase transitions, respectively. The
diagram has of course mirror symmetry with respect to the $I/J $ axis.

\end{document}